\documentclass[onecolumn]{aa}

\usepackage{graphicx}
\usepackage{txfonts}
\begin{document}

   \title{Discovery of an unusual  bright  eclipsing binary with the longest known period:  TYC 2505-672-1 / MASTER OT J095310.04+335352.8}
   \author{V. Lipunov  \inst{1} \inst{2}
          \and E. Gorbovskoy \inst{2}
          \and V. Afanasiev \inst{3}
\and A. Tatarnikova\inst{2}
          \and D. Denisenko \inst{2}
          \and D. Makarov \inst{3}
          \and N. Tiurina \inst{2}
          \and V. Krushinsky\inst{4}
\and A. Vinokurov\inst{3}
          \and P. Balanutsa\inst{2}
          \and A. Kuznetsov \inst{2}
          \and O. Gress \inst{5}
          \and Yu.Sergienko \inst{6}
          \and V. Yurkov \inst{6}
          \and A. Gabovich \inst{6}
          \and A. Tlatov  \inst{7}
          \and V. Senik \inst{7}
          \and V. Vladimirov \inst{2}
          \and E.Popova \inst{2}
          }
   \institute{ Lomonosov Moscow State University, Physics Department, Leninskie Gory, 1, Moscow 119991 Russia\\
              \email{lipunov2007@gmail.com}
         \and Lomonosov Moscow State University, Sternberg Astronomical Institute, Universitetsky prospekt, 13, Moscow, 119234, Russia\\
\and Special Astrophysical Observatory, Nizhnij Arkhyz, Zelenchukskiy reg., Karachai-Cherkessian Republic, 369167, Russia  \\
             \and Kourovka Astronomical Observatory, Ural Federal University, Lenin ave. 51, Ekaterinburg 620000, Russia\\
             \and Applied Physics Institute, Irkutsk State University, 20, Gagarin blvd,664003, Irkutsk, Russia\\
\and Blagoveschensk State Pedagogical University, Lenin str., 104, Blagoveschensk, Amur region, 675000, Russia\\
              \and Kislovodsk Mountain Astronomical Station of the Pulkovo observator, Gagarin str., 100, Kislovodsk, 357700, Russia}
   \date{Received 2015; accepted 2015 }
\abstract{
   We report on the MASTER Global Robotic Net discovery of an eclipsing binary, MASTER OT J095310.04+335352.8, previously known  as  unremarkable star TYC 2505-672-1, which displays extreme orbital  parameters. The orbital period P=69.1 yr is more than 2.5 times longer than that of $\varepsilon$-Aurigae, which is the previous record holder. The light curve is characterized by an extremely deep total eclipse with a depth of more than 4.5 mag, which is symmetrically shaped and has a total duration of 3.5 yrs. The eclipse is essentially gray. The spectra acquired with the Russian 6 m BTA telescope both at minimum and maximum light mainly correspond to an M0-1III--type red giant, but the spectra taken at the bottom of eclipse show small traces of a sufficiently hot source. The observed properties of this system can be better explained as the red giant eclipsed by a large cloud (the disk) of small particles  surrounding the invisible secondary companion.}

 \maketitle
%
\section{Introduction}
 Very wide binaries with very long orbital periods are  important objects of modern stellar astronomy because the distribution of their separations is a clue to perturbations in the gravitational field of the Galaxy ~\citep{Jiang}.
There are only seven eclipsing variable stars with known orbital periods longer than ten years in the last edition of the  General Catalogue of Variable Stars ~\citep{samus07,samusvizier}. The most famous of these stars is $\varepsilon$-Auriga, which has the longest known orbital period (27.1 yr).

TYC 2505-672-1 has never been a popular target for observers, but, as
we demonstrate below,  its orbital period is more than 2.5 times longer than that of $\varepsilon$-Auriga. The spectral type of TYC 2505-672-1 was reported  to be M2III and its estimated distance was reported to be near 1179 pc ~\citep{pickles}. Combined with the UCAC4 proper motion (3.5, -7.4) mas/yr ~\citep{zacharias} this corresponds to a tangential velocity of 46 km/s. The Galactic coordinates of the source are (191.7, +51.3) and extinction in its direction is negligible.

During the usual night sky survey, the MASTER-Amur auto-detection system ~\citep{Lipunov10,Lipunov07,Gor10} originally discovered the deep fading (the depth is greater than 4.5 mag) of the bright star TYC 2505-672-1, whose variability was previously unknown.
MASTER-Amur is a node of MASTER (Mobile Astronomical System of the TElescope Robots) Global Robotic Net.

The optical antitransient  MASTER OT J095310.04+335352.8 was   initially  bright, with an unfiltered magnitude of about 10.5mag, on the MASTER-Net archive images from 2009-12-01.007 UT in MASTER-Kislovodsk, 2010-03-10.587 UT in MASTER-Tunka, and  2009-11-17.775 UT in the MASTER-Amur database. In November 2011 - March 2012, however, after the seasonal gap in observations due to the conjunction with the Sun, the star was found to fade from R=13.0 to R=14.1. Since then it has resided at a low brightness state for almost a year.
There have only been several such long-period sources known before, and so we began a photometrical and spectral  analysis of this object.

The aim of this paper is to present our analysis of an extensive collection of archival and new photometrical data of  MASTER OT J095310.04+335352.8 to define its type of variability and to estimate physical properties of the object on the base of our spectral observations. We also  compare the spectral and photometrical  behavior  of TYC 2505-672-1 with that of  $\varepsilon$-Auriga and consider some similar points in these systems.

\section{Observations}

\subsection{MASTER discovery }

  The optical source MASTER OT J095310.04+335352.8, also known as unremarkable star TYC 2505-672-1, was originally discovered by the MASTER Global Robotic Net  ~\citep{Lipunov10,kornilov,Lipunov07,Gor13,Lipunov04} auto-detection system  at (RA, Dec) = $09h 53m 10s.00$, Decl. = $+33^\circ$ 53$'$  52.7$''$  as an optical antitransient ~\citep{d13}; see Figure 1.

  The MASTER Global Robotic Net  consists of six identical MASTER-II observatories. Each observatory has twin 0.40-m f/2.5 reflectors + 4Kx4K CCD, which provide an  8-square-degree field of view. The  observatories are as follows: operating in Russia, from east to west, the MASTER-Amur, MASTER-Tunka, MASTER-Ural, and MASTER-Kislovodsk;    in South Africa, the MASTER-SAAO; in the Canary Islands, the MASTER-IAC.
In addition, the very wide-field camera, MASTER-VWFC, in Argentina has an 800-square-degree field of view (MASTER-II-Argentina is currently in development).
     The typical limiting magnitudes on the MASTER-II instrument on the single images taken in survey mode range from 19-20 mag (60 s exposures) to 20-21.5 mag (180 s images). The primary goals of the MASTER Global Robotic Net are the rapid response to gamma-ray bursts (GRB) in alert mode and the discovery of optical transients in survey mode. All images are reduced  automatically in real-time mode via our  software developed by the MASTER team over the past ten years. This fully automatic software has to date discovered more than 1000 optical transients; see $ http://observ.pereplet.ru/MASTER\_OT.html $.

There is the short description of MASTER reduction software, which allows us to discover any transients in the widefield images ($2^\circ x 2^\circ$) in an online regime. An object is detected as a transient if it appears in  two images  in the same night and if there is no match within a $5"$ circle in either the  USNO-B catalog or the MASTER reference images.
The formal criterion for optical transient detection is a change in the magnitude by more than 2 from its catalog value. Most of the MASTER survey images are obtained in white (unfiltered) light to increase the limiting magnitude. The corresponding internal photometric magnitudes can be described fairly well by the equation $W=0.8*R2+0.2*B2$, where $R2$ and $B2$ are the second epoch DSS red and blue magnitudes, respectively, adopted from USNO-B1.0 catalog ~\citep{monet}.
The absolute majority of MASTER optical transients (OTs) are  flares of various types (optical counterparts of GRBs, supernovae, QSO flares, dwarf novae, novae, nova-like, UV Cet type flare stars, etc).

    The unusual MASTER OT J095310.04+335352.8 object has met the optical transient criterion on the MASTER-Amur and MASTER-Kislovodsk images acquired in the end of 2012 to the beginning of 2013. The previously unremarkable star TYC 2505-672-1 (R.A. = $09h 53m 10s.00$, Decl. = $+33^\circ$ 53$'$  52.7$''$, V=10.71, B=12.51) = 2MASS J09531000+3353527 (J=7.61, H=6.78, K=6.57) faded by $\sim$ 4 mag from its ordinary level. Figure 1 compares MASTER-Amur images acquired in 2011 (maximum) and 2013 during a deep dimming event. Table 1 lists selected photometry obtained by MASTER telescopes.

\begin{table}
      \caption{ MASTER photometry of the MASTER OT J095310.04+335352.8 in 2011-2013}
         \label{table:1}
\begin{tabular}{c c c}
\hline
 Observational Date,UT & Mag,filter& Observatory \\
\hline
2011-02-03 15:15:47  & 10.34W  & MASTER-Amur \\
2011-03-14 10:43:18  &  10.03V &        MASTER-Amur \\
2011-04-07 12:30:22  &  10.44W  &       MASTER-Amur \\
seasonal gap& due to solar& conjunction \\
2011-11-03 23:44:28  & 12.62W  & MASTER-Kislovodsk \\
2011-11-04 00:17:08  & 12.64W  & MASTER-Kislovodsk \\
2011-11-09 17:42:27  & 12.99R   & MASTER-Amur \\
2011-11-20 16:58:19  & 13.10R  & MASTER-Amur \\
2011-11-29 16:47:27  &  13.27R  &       MASTER-Amur \\
2011-12-09 16:14:25  &  13.44R  &       MASTER-Amur \\
2012-01-08 13:37:51  &  13.85R &        MASTER-Amur \\
2012-03-01 12:50:18  &  14.06R  &       MASTER-Amur \\
2012-03-17 14:40:17  &  14.31V  &       MASTER-Amur \\
2012-04-02 18:09:36  &  13.81W  &       MASTER-Amur \\
seasonal gap& due to solar& conjunction \\
2012-01-08 13:37:51  &  13.85R  &       MASTER-Amur\\
2012-03-01 12:50:18  &  14.06R  &       MASTER-Amur\\
2012-03-17 14:40:17  &  14.31V  &       MASTER-Amur\\
2012-04-02 18:09:36  &  13.81W  &       MASTER-Amur\\
seasonal gap& due to solar& conjunction \\
2012-09-29 20:02:30  &  13.98W  &       MASTER-Amur\\
2012-10-25 19:23:53  &  14.03W  &       MASTER-Amur\\
2012-12-19 15:46:57  &  14.01W  &       MASTER-Amur\\
2013-01-07 16:46:19  &  13.77W  &       MASTER-Amur\\
2013-02-01 16:51:19  &  13.82W  &       MASTER-Amur\\
\hline
\end{tabular}
\end{table}

\subsection{ Spectroscopy and photometry}

All spectra of  MASTER OT J095310.04+335352.8 were obtained with the Russian 6 m BTA telescope (SAO RAS, Russia) + SCORPIO instrument ~\citep{va05,va11}.  The first spectrum was obtained during an eclipse on February 10, 2013 (see fig. 2 and ~\citep{va13}). The spectrum in the red range was consistent with that of a late giant without unusual absorption lines, but with $H\alpha$ in emission $(FWHM=320km/s)$. The observed wavelength of $H\alpha$ line was $6561.0\pm 0.1 $ {\AA} corresponding to the blue shift of $-82 km/s$. A number of weak FeI and NaI emission lines were also present with typical FWHM of $\sim 150-180 km/s$.

The second and third spectra were obtained after  eclipse (October 26, 2014, a blue spectrum; March 19, 2015, a red one). The last spectrum is shown in Fig. 3. It is a pure late-type giant spectrum. We note that H$\alpha$ and the NaI doublet are in absorption although they were in emission during eclipse. The emission components of the lines may possibly also be present, but their intensities are near noise level because of the very strong spectrum of the red giant. The blue spectrum (October 26, 2014) and spectral energy distribution (SED) of a standard M1 giant (see spectral catalog of  ~\citep{Silva} are shown in Fig. 4. The spectral energy distribution of the hot components are shown in fig. 5. We tried to approximate the observed hot component spectrum by means of SEDs for MK standard stars for spectral classes from O to G. The best result was shown in Fig. 5. The curves coincide poorly, which may be due to the small wavelength interval covered by this SED and the poor signal-to-noise ratio in the blue part of the spectrum (more details are given in Sect. 4).

Following the MASTER discovery announcement, the new variable star was initially added to the AAVSO international database and Variable Star Index ~\citep{watson}  as a suspected R CrB star. The behavior of the object was  initially puzzling. It was supposed to be a new R CrB type variable (the red supergiant ejecting carbon clouds at irregular intervals of time). The AAVSO assumed that this source was a new R CrB-type variable since observations showed that the red supergiant ejected carbon clouds at irregular intervals of time.
 However, the typical dimming phase of R CrB stars lasts from one week to one month, whereas  TYC 2505-672-1 took more than three months to fade from 10 mag   to  14 mag. The star's brightness was nearly constant in 1999-2000 during  observations by the NSVS (ROTSE-I) project ~\citep{carl2000,carl2}.

Previous fadings were not detected either on the nine Palomar plates taken from 1955 to 1996 or on the 31 Palomar/NEAT images taken during nine different nights from 2001-2003. The object was not detected in 1RXS and IRAS bright source catalogs. Follow-up spectroscopic and photometric observations of the TYC 2505-672-1 and archival searches for historical fading episodes were requested in ATel 4784.  As a result, the recovery from the deep fading in 2013-2014 was traced very well  by several optical observers  using CCDs with BVRI filters.
 The data  show a gradual of flux increase in all photometric bands without a color change (gray eclipse) except in the $B$ band. This behavior is not typical for R CrB stars.  As of October, 2014, after the yearly gap due to the solar conjunction, the star has returned to its pre-eclipse level of V=10.8 mag, B=12.5 mag.

\subsection{ Historical eclipse}

On June 23, 2014 the third release of the digitized Harvard College Observatory plates ~\citep{grin} was published covering Galactic latitudes from +60 to +45. R. Jansen ~\citep{jans}  checked the position of MASTER OT J095310.04+335352.8 on the DASCH project website $ http://dasch.rc.fas.harvard.edu/ $ and found an additional eclipse of this star in 1942-1945. The comparison of DASCH images from 1968 (maximum) and 1945 (fading) is shown in Fig. 6. The DASCH observations of this field cover the time interval from 1890 to 1990 with one significant gap in 1953-1967. Fig. 7 shows the light curve of  MASTER OT J095310.04+335352.8 = TYC 2505-672-1 comprised of MASTER, Harvard plates, NSVS, and Catalina Sky Survey ~\citep{drake}  observations. Magnitudes measured in various photometric bands were reduced to the common zero point using different color adjustments. For example, B-band photographic magnitudes from Harvard plates were corrected by -1.8 mag and unfiltered (W) MASTER observations were corrected by +1.45 mag, respectively.
Thus, the mystery of this object was solved. MASTER OT J095310.04+335352.8 turned out to be a deeply eclipsing binary star with an unprecedentedly  long interval between  eclipses.

\section{ Parameters of the binary system}

Using the combined data from 1890 to 2014, we determined the orbital period of this new variable to be  P=25245, which is equal to 69.1 yr with a 0.1 year accuracy; this accuracy is supported
by the end  of two eclipses. We rule out the possibility of a fractional value of this period. Taking the gap in 1965-1967 data and a lack of accurate data in 1990-1992 into account, the P/3 value is  23 years. However, the Harvard data in 1920-1922 and 1897-1899 show no fading of this star 23 and 46 years before the 1942-1945 eclipse, thereby ruling out this possible alias.

Figure 7 shows the folded light curve with the 69.1 yr period. Thus MASTER OT J095310.04+335352.8 becomes the eclipsing variability with a period at least 2.5 times longer than those of all previously known objects with measured periods.
These periods are listed in Table 2 with their types, periods in days and years for convenience, and magnitude ranges.

The second column corresponds to the eclipsing variable of Algol type (EA in AAVSO clasification);
 the GS subtype indicates  a  system with one or both  giant and supergiant components; the SRC refers to semiregular pulsations of a late-type supergiant. A colon (:) means that the listed value is uncertain.

\begin{table}
      \caption{ Previously known eclipsing variables with periods longer than ten years (3652.5 days)}
         \label{table:2}
          \centering                                      
\begin{tabular}{c c c c c}
\hline 
 Star name& Variability type & Period,d & Period,y  & Range,mag \\
\hline
V0695 Cyg       &EA/GS/D        &3784.3 &10.36  &3.73-3.89V\\
V0383 Sco&      EA/GS   &4875.9 &13.35  &10.6-16.3V\\
&       +SRC    &       &       &\\
$\gamma$ Per    &EA/GS  &5346   &14.64  &2.91-3.21V\\
V0381 Sco       &EA/GS  &6545   &17.92  &11.63V-16.0P\\
        &+SRC   &       &       &\\
VV Cep  &EA/GS  &7430   &20.34  &4.8-5.36V\\
        &+SRC   &       &       &\\
NSV6116 &EA:    &9443   &25.85  &4.32-?V\\
($\alpha Com$)  &       &       &       & \\
Eps Aur &EA/GS  &9892   &27.08  &2.92-3.83 V\\
\hline
\end{tabular}
\end{table}

The orbital period of NSV 6116 $(\alpha  Com)$, which is a visual and double-lined spectroscopic binary with an inclination very close to $90^{\circ}$,  is determined from long-term optical observations. The initial calculation by ~\citep{mut10} showed the eclipses of $\alpha Com$ to be highly likely, with a predicted closest projected approach around 2015 Jan. 24 using the period value of 9485.7d. As the event approached, the new findings showed that the prediction was in error. The revised value of the period is $9443.1\pm 3.0$ days  ~\citep{mut15}. The eclipse  likely occurred two months earlier than predicted, around the 20th of Nov 2014, and has not been observed.
Orbital periods of these seven stars and of MASTER OT J095310.04+335352.8 are plotted in Fig. 8. This new MASTER variable star stands out in this plot, so it is clear that  its period is exceptionally long.

The mass of the eclipsing body is unknown since it is entirely hidden by the surrounding dust cloud. We assume for  simplicity of calculation that the total mass of the system is about $4 M_\odot$. From the generalized Keplers third  law, we then obtain the semimajor axis of about 26.7 AU and the orbital velocity of 11.55 km / s (3.0 AU / yr). The eclipse duration is 3.5 yr (0.05 P) with the ingress lasting 1 yr, total phase 1.5 yr, and egress again 1 yr, hence,  the total size of the eclipsing body is about 7.5 AU (i.e., 1 billion km, $10^{14}$ cm or 750 solar diameters). The nominal size of the M1III-type red giant is 60 times that of the Sun (0.6 AU in diameter). Since  partial phases of the eclipse last about a year rather than 0.2 yr, the eclipsing body should have its density increasing toward the center. The height (thickness) of the eclipsing body should exceed the diameter of the primary M1III-type star (0.6 AU). So, the vertical to horizontal size ratio may be as small as 1:10 for edge-on geometry.
Hence, as the eclipse is achromatic, we can only tell about the gray  absorbtion by the big size particles.

 The main arguments   for edge on is the  absence  of rebrightening   during eclipse phase, which we can see in the  $\varepsilon$-Aur case. This means that we do not see the internal hole in the disk and  center of a hot star, as one  can see in $\varepsilon$-Aur and that allows us to see the rebrightening of its system.
Therefore, as the dust disks are very cold, in that they have a low velocity  dispersion, they must be very thin. Taking into account that we do not see the rebrightening and that the disk must be thin (from the low dispersion), we find that the  disk's plane is close to the line of sight, i.e., very close to edge-on. This can be not the only scenario, of course, and further study of this system can provide astronomers with possibility of another point of view to explain the MASTER OT J095310.04+335352.8 structure.

\section{Discussion}

Considering the above data,  MASTER OT J095310.04+335352.8 is similar to the famous eclipsing binary $\varepsilon$-Auriga, but this source is more extreme and is different from the last source.

First, the eclipse depth in  MASTER OT J095310.04+335352.8 is five times deeper than in $\varepsilon$-Auriga, despite the fact that the central star is hotter and more luminous (F0I). This  difference may be explained by the different viewing geometry of the two binary systems.  Since the inclination of the orbital plane of MASTER OT J095310.04+335352.8 is closer to $90\deg,$ we see a deeper eclipse. We observe this deeper eclipse taking into account that 1) the absence of a rebrightening inside the eclipse period in which observer would  see the internal central disk's hole with its central star, which is visible in  $\varepsilon$-Auriga; and 2) the dust disk is a cold disk with low dispersion, i.e., the thin disk, and the central star is smaller ($\sim60 R_{\bigodot} \div \sim 200 R_{\bigodot}$).

The second difference is the eclipse shape. In MASTER OT J095310.04+335352.8, the eclipse is symmetrical with a flat bottom, whereas in $\varepsilon$-Auriga the ingress is
shallower and egress is steeper ~\citep{stencel08}.

Third, ~\citep{maslov} noted that $\varepsilon$-Auriga has shallow secondary minimum on IR light curves.
Finally, there is a rebrightening in the middle of the eclipse in $\varepsilon$-Auriga, which is explained by the gap or a hole in the center of the occulting disk that is supposedly swept by the binary star inside ~\citep{stencel12}. According to the recent infrared observations with Herschel Space Observatory ~\citep{hoard}, the cool  dust disk in $\varepsilon$-Auriga has an overall temperature of about 550 K, and the side of the disk facing the bright F-type star is warmer (1150 K).

There are two systems in Table 2 that have an eclipse depth of about five magnitudes, which is similar to MASTER OT J095310.04+335352.8. However, these two stars (V381 Sco and V383 Sco) are EA/GS+SRC type variables. In these systems, semiregular pulsating stars (red supergiants) are eclipsing the hotter components. According to ~\citep{galan}, the V383 Sco system consists of a pulsating M-type supergiant that periodically obscures the much more luminous F0I star. The eclipses in V383 Sco have different depth in various photometric bands, with the color index (V-I) changing by more than 1.5 mag between the maximum and minimum light.

 The principal difference between MASTER OT J095310.04+335352.8 and such systems is that the obscuring body of MASTER OT J095310.04+335352.8 cannot be self-luminous. As in the case of $\varepsilon$-Auriga, we are dealing with a dust disk that absorbs the light of the central star(s) resulting in the gray eclipse. Also, no significant pulsations in maximum light are present in the historical light curve of MASTER OT J095310.04+335352.8, which also speaks in favor of the cold dust disk.
Of course, the edge-on disk orientation is not the only possibility to explain this system, and further study of this system will clarify this question.

The spectral energy distribution (SED) of the system was very similar in the all our spectra despite the great difference in continuum fluxes.
The main contributor to the SED is a red giant in
eclipse as well as after eclipse, and the TiO bands of the red giant are the most prominent absorption features in the spectra. The spectrum obtained  in eclipse also exhibits some faint emission lines against the TiO bands (see Sect. 2.2). We neglect the interstellar reddening because the system is  above the Galactic plane and there is not any sufficient continuum distortion.

The spectrum obtained in eclipse (February 10, 2013) includes the red region, allowing us to estimate the spectral type of the partially eclipsed red giant by means of [TiO]$_1$ spectral index  ~\citep{kenyon}. The star is an M1-type giant ($\pm 0.5$ subclass). Differences between the observed spectrum and standard SEDs of M0III and M1III do not exceed $10\%$ in the red part of spectrum, but for M2III the difference are about $20\%$. There are not any signs of CaH absorption bands in any of our spectra, therefore, the luminosity class of the star cannot be V (MS star). Another luminosity criterion is a relative intensity of    $FeI 5250$ {\AA} and $CaI 5269$ {\AA}. According to this criterion the star is a giant (III class) rather than supergiant (I class).  We note that the SED in the red region is nearly the same as that of a standard M1-type giant, but in the blue region the situation is quite different; see Figs. 2 and  5. The continuum fluxes are nearly 20 $\%$ higher than the standard M1III SED.

The second spectrum was obtained after eclipse and does not include the red region and, therefore, we cannot trace the emission lines evolution. The main difference between  two spectra is that  after eclipse the observed SED is similar to a standard M1 SED even in the blue region; see Fig.4.

A difference between the SED observed in eclipse and a standard M1III SED is shown in  Fig. 5.  A SED of the difference characterizes  a sufficiently hot source because its continuum flux grows from the visual to blue region. The emission of this source was visible only during eclipse (when the contribution of the red giant to the total flux was greatly diminished). Without the eclipse, the SED is the pure contribution from the M1III-type star. The SED of the hot component corresponds poorly to a SED of any standard hot star with spectral class earlier than F5  and does not correspond to SEDs of later stars (thick line in Fig.5 is a standard B6V SED); we tried to approximate the observed hot component
spectrum by means of SEDs for MK standard stars for spectral classes from
O to G. The fact that the situation is less than ideal may be due to

 the small wavelength interval covered by this SED and the poor signal-to-noise ratio in the blue part of the spectrum. Numerical estimates of the hot component's  temperature or luminosity cannot be made   without observations in the UV range.

There is one more confirmation of the faint hot component appearance during eclipse. According to AAVSO data ~\citep{kafka}, the color index B-V was bluer near the center of the eclipse (B-V=1.0-1.2 mag) than during egress and after eclipse when the B-V color index was nearly constant (B-V=1.6 mag).

A comparison of TYC 2505-672-1 with $\varepsilon$-Auriga (another long period binary with similar features in its light curves) leads us to conclude that both systems have common features in their SEDs. Although in the case of $\varepsilon$-Auriga system we observe an F2-type supergiant instead of an M1-type giant, both systems demonstrate faint traces of a hot component (imbedded in the center of a huge dust disk). The existence of the hot component (spectral class  B1) for $\varepsilon$-Auriga was proved by analyzing the UV spectra obtained by International Ultraviolet Explorer.

High-resolution spectroscopy ($\lambda/\bigtriangleup\lambda\sim 20000$) is needed to determine the component mass ratio of this unique binary system. Owing to the exceptionally long orbital period, the changes of radial velocity will become noticeable in about eight years (P/8) after  mid-eclipse time. Precision photometric observations will also be useful around the orbital phase 0.5 (expected 2047) to search for possible reflection effects. Far-infrared observations are encouraged to determine the spectral contribution of the dust disk and the size of its constituent particles. Also, suspected variable stars, especially M-type giants with RCB, and E: classification
in the NSV catalog should be checked in databases such as DASCH for  past eclipses.
In any case, the main difference of MASTER OT J095310.04+335352.8 from $\varepsilon$-Auriga is its position with respect to Galactic plane.

The M1 giant at 50 degrees above the plane is located on 1kpc from the Galactic plane. This means that it is spaced between the old stars of the disk.

 And there is some problem with the existence of an extended dust disk that lives much more than $\sim 10^7$ years and is in the system inside old star's galactic disk. On the other  hand, we can tell whether  a huge dust disk  is fed by dust accretion from the giant.

In this case, the posibility of blue color excess emergence at the time of  eclipse becomes clear.

 It is obvious that new observational data will clarify the discussion.

\section{CONCLUSION}

We have discovered an eclipsing binary system with extreme properties: an unprecedentedly  long period of 69.1 yr and a very deep eclipse reaching at least 4.5mag in the V band, explained by a cold dust disk. This discovery has been made  possible by combining the unique search capacities of the modern MASTER robotic telescope system and the  digitization of  archival photographic plates that cover more than a century of observations. Our finding shows once again that the sky is full of extraordinary objects awaiting  identification. Combination of  state-of-art survey projects and data mining of the numerous archives  at astronomical observatories worldwide can bring many unexpected discoveries, even among bright objects.

\begin{acknowledgements}

MASTER project is supported in part by the Program of Development of Lomonosov Moscow State University. This work was partially supported by RFBR 15-02-07875. We are thankful to several AAVSO observers who have reported their BVRI photometry of MASTER OT J095310.04+335352.8 to AAVSO international database.

The authors thank Joao Alves  for important remarks concerning high latitude position of our object. We thank Prof.Tony Rothman for a fruitful discussion.
\end{acknowledgements}

\newpage
   \begin{figure}
   \includegraphics[width=84mm]{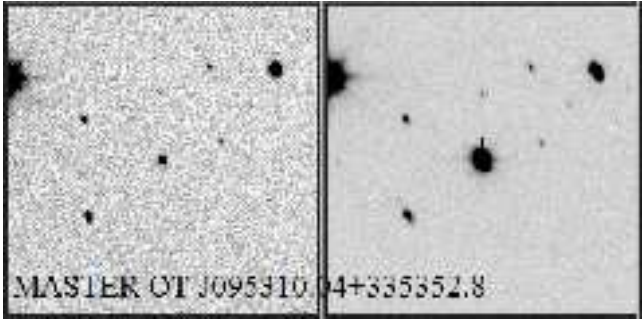}
      \caption{ MASTER-Amur images of MASTER OT J095310.04+335352.8 = TYC 2505-672-1 in deep fading (2013; left panel) and in maximum (2011; right panel)}
         \label{Figure1}
   \end{figure}

\newpage
\begin{figure}
   \includegraphics[width=84mm]{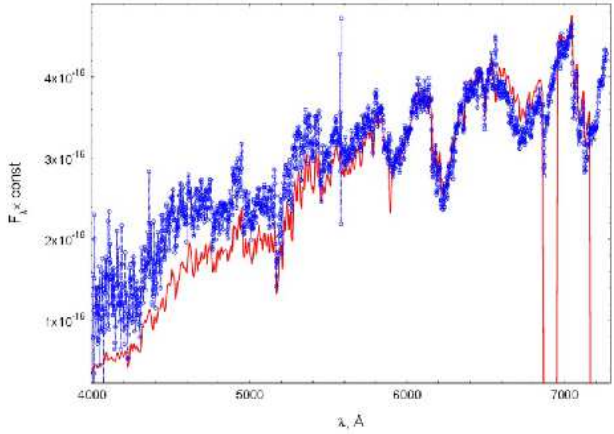}
      \caption{ Spectrum of  MASTER OT J095310.04+335352.8, obtained during eclipse on February 10, 2013 (points connected with the thin line) and spectral energy distribution of a standard M1-type giant (zero fluxes in the model energy distributions indicate the spectral regions with strong telluric absorption bands). }
         \label{Figure2}
   \end{figure}
\newpage
\begin{figure}
   \includegraphics[width=84mm]{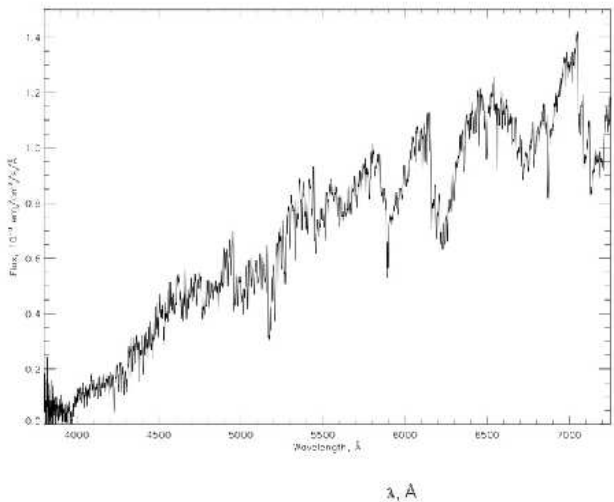}
      \caption{ MASTER OT J095310.04+335352.8 spectrum obtained after  eclipse on 2015-03-19.  One can see the red part of an after eclipse spectrum. We note that $H\alpha$ and $NaI$ doublet are in absorption although they were in emission during eclipse. Emission components of the lines may also be present, but their intensities are near noise level because of the very strong spectrum of the red giant.}
         \label{Figure3}
   \end{figure}
\newpage
\begin{figure}
   \includegraphics[width=84mm]{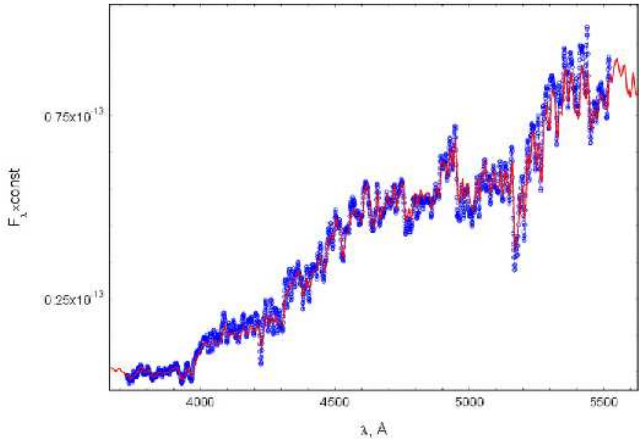}
      \caption{ MASTER OT J095310.04+335352.8 spectrum obtained after  eclipse on October 26, 2014
(points connected with thin line) and spectral energy distribution of a standard M1 giant (thick line).}
         \label{Figure4}
   \end{figure}
\newpage
   \begin{figure}
   \includegraphics[width=84mm]{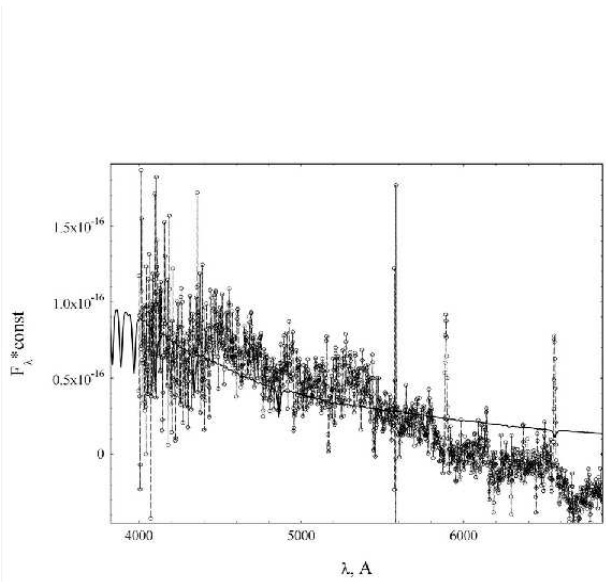}
      \caption{ Spectral energy distributions of the "hot component" (open circles) and a standard B6V star (thick line). }
         \label{Figure5}
   \end{figure}
\newpage
   \begin{figure}
   \includegraphics[width=84mm]{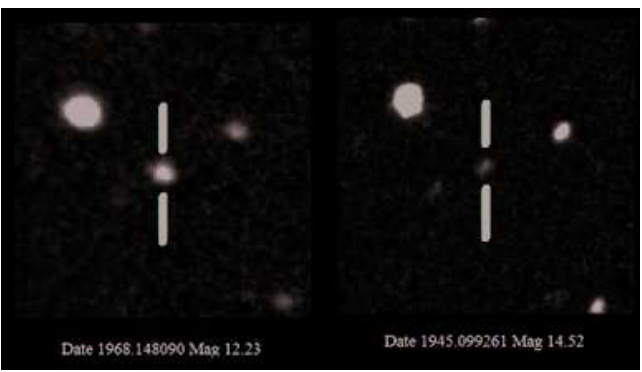}
      \caption{ DASCH images of TYC 2505-672-1 at the maximum light (1968) and in eclipse (1945).}
         \label{Figure6}
   \end{figure}
\newpage
   \begin{figure}
   \includegraphics[width=84mm]{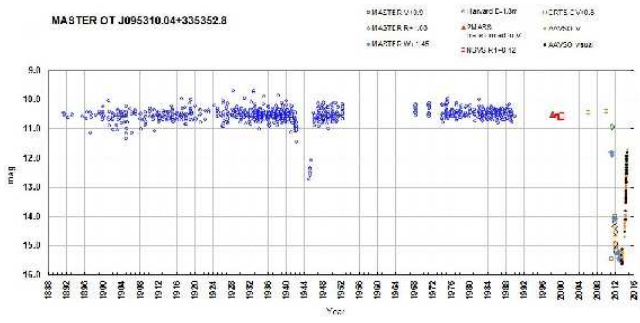}
      \caption{ Historical light curve of MASTER OT J095310.04+335352.8 = TYC 2505-672-1 combined from various sources with the corresponding color corrections applied.}
         \label{Figure7}
   \end{figure}
\newpage
   \begin{figure}
   \includegraphics[width=84mm]{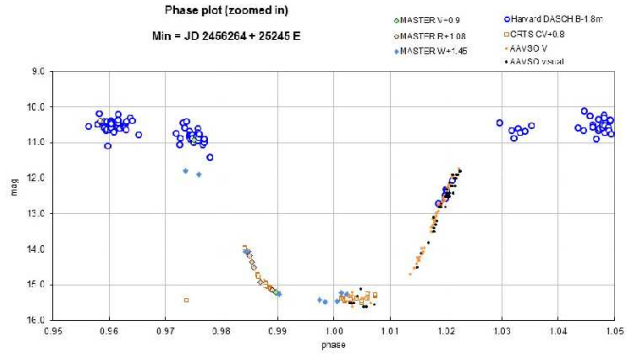}
      \caption{ Light curve of MASTER OT J095310.04+335352.8 folded with the 69.1 yr period. Phase interval 0.95-1.05 is zoomed in at the bottom showing the symmetric shape of the eclipse.}
         \label{Figure8}
   \end{figure}


\end{document}